\documentclass[journal]{IEEEtran}

\usepackage{cite}

\usepackage{graphicx}
                        
\usepackage{hyperref}
\usepackage{amsmath}

\begin{document}


\title{Pump-probe differencing technique for cavity-enhanced, noise-canceling saturation laser spectroscopy}

\author{Glenn de Vine,
        David E. McClelland,
        John D. Close
        and~Malcolm~B.~Gray 
\thanks{Centre for Gravitational Physics, Faculty of Science, The Australian National University, Canberra, ACT 0200, Australia.  John Close is with Centre for Quantum Atom Optics, Faculty of Science, The Australian National University, Canberra, ACT 0200, Australia}}

\maketitle

\begin{abstract}We present an experimental technique enabling mechanical-noise free, cavity-enhanced frequency measurements of an atomic transition and its hyperfine structure. We employ the 532nm frequency doubled output from a Nd:YAG laser and an iodine vapour cell. The cell is placed in a traveling-wave Fabry-Perot interferometer (FPI) with counter-propagating pump and probe beams. The FPI is locked using the Pound-Drever-Hall (PDH) technique. Mechanical noise is rejected by differencing pump and probe signals. In addition, this differenced error signal gives a sensitive measure of differential non-linearity within the FPI.\end{abstract} 



Fabry-Perot interferometers provide an obvious means of amplifying the response of weak atomic transitions \cite{CerezIEEE80}.  The interferometer bounce number yields the effective increase in interaction length given by the FPI \cite{regehr}.  However, this interferometric improvement necessarily couples laser frequency noise and mechanical noise of the interferometer into the measurement.  This places stringent requirements on the free running frequency noise of the interrogating laser as well as the acoustic and seismic isolation required for interferometer operation. 

To address these issues the NICE-OHMS \cite{YeJOSAB98} technique provides a cavity enhanced technique that gives immunity to both laser frequency noise and interferometer mechanical noise whilst realising cavity enhanced sensitivity.  NICE-OHMS achieves this by using phase modulation (PM) at the free spectral range (FSR) of the interferometer in order to probe the refractive index changes around the FPI resonance experienced by the laser carrier electric field.  The PM sidebands resonate in adjacent FPI modes and experience an identical response to both frequency noise of the laser and cavity mechanical noise as does the carrier (coming from the same laser and interrogating the same cavity).  The identical response preserves PM symmetry and provides noise immunity. A second, non-resonant PM is required to lock the laser carrier to a FPI resonance whilst dual FM demodulation is required to lock the RF source dynamically to the FPI FSR.  A third low frequency modulation is applied to the FPI in order to give a signal proportional to the derivative signal, yielding a noise immune output proportional to the atomic transition being measured. 

Here we introduce a new technique that provides both cavity enhanced sensitivity and noise immunity in a far simpler experimental configuration.  Furthermore, our technique also provides a new and sensitive measure of the differential non-linearity of low loss optical media.

\begin{figure}[htb]
  \begin{center}
  \includegraphics[width=8.5cm]{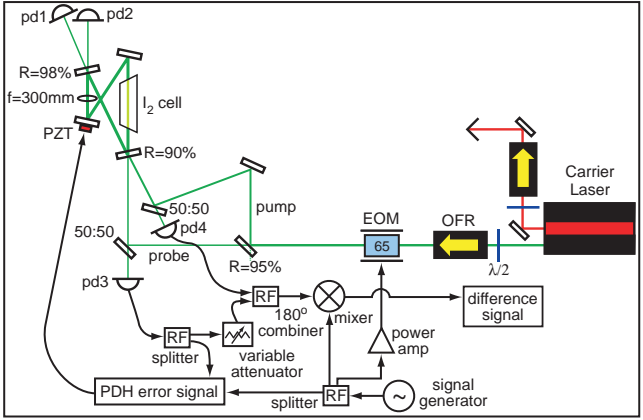}
  \end{center}
  \caption[TESTX]{Schematic layout of experiment. $\lambda/2$ half waveplate; OFR optical Faraday rotator (optical isolator); EOM electro-optic modulator; pd photo-detector.}
  \label{fig1}
\end{figure}

Our cavity-enhanced technique uses a travelling-wave FPI which is interrogated in both directions (see figure \ref{fig1}). The laser beam, prior to pump/probe split off, is phase modulated at a high RF frequency.  This ensures that the phase modulation is identical for both the pump and the probe and any residual amplitude modulation (AM) caused by the modulator appears equally on both beams.  The pump beam interrogates the cavity and the reflected beam is incident on an RF photodetector prior to demodulation.  This yields the PDH \cite{pdh} error signal for the pump mode relative to the laser frequency.  The probe beam is also reflected onto itÕs own RF photodetector.  The RF signals from the pump and probe are then differenced on a 180$^{o}$ RF splitter/combiner prior to demodulation.  By balancing the RF signal power of the pump relative to the probe, prior to subtraction and demodulation, it is possible to cancel out the error signal resulting from cavity detuning, including mechanical FPI noise, as this is common to both. 

The demodulated PDH pump error signal is used to lock the ring cavity to the laser frequency while the demodulated difference output then yields the differential phase response between the pump and probe beams.  As the probe beam is far weaker than the pump beam, the shot noise limit of the subtracted output is effectively given by the probe sensitivity alone.  Day et al \cite{day} calculate the shot noise limited frequency sensitivity of the PDH error signal as:

\begin{eqnarray}
 S  \approx  \delta \nu_c \sqrt{\frac{h\nu}{8P_{in}\eta}}~~(Hz/\sqrt{Hz})   \label{S}
\end{eqnarray}

Where Planck's constant, $h = 6.626\times10^{-34} Js$. Using an incident probe power, P$_{in}$, of 500 $\mu watts$, a cavity linewidth, $\delta \nu_c$, of 8.5 $MHz$ (a cavity Finesse, $\mathcal{F}$, of 50), and a photodetector responsivity, $\eta$, of 0.4 Amperes/Watt at a wavelength, $\lambda$, of 532nm gives a shot noise limited frequency sensitivity of 0.13 $Hz/\sqrt{Hz}$.

This can readily be modified to yield the shot noise limited sensitivity to intra-cavity phase signals caused by an atomic transition:

\begin{eqnarray}
  \phi _{shot} = \frac{\pi}{\mathcal{F}}\sqrt{\frac{h\nu}{2P_{in}\eta}}~~(rad/\sqrt{Hz})   \label{phishot}
\end{eqnarray}

Using the same interrogation details as above, this yields a shot noise limited phase sensitivity of 1.25 $nanoradians/\sqrt{Hz}$. Due to the noise immunity properties of both the NICE-OHMS technique and the technique presented here, this shot-noise limited sensitivity is experimentally achievable.

Figure \ref{fig1} shows a simplified schematic of our experimental set up.  We use the doubled output of a Prometheus (INNOLIGHT GmbH) laser ($\approx$20mW at 532nm) to provide both the pump and probe beams for our experiment.  After optical isolation, the laser output is phase modulated at 65MHz (NewFocus 4002).  The mode matching optics are then traversed prior to splitting the pump (8.7mW) and probe (0.45mW) beams.  The pump and probe beams then interrogate the ring cavity in different directions.  The FPI consists of a $7\%$ input coupler, two high reflector mirrors and a $2\%$ output coupler giving a nominal finesse of $\approx$67, ignoring intra cavity losses.  The reflected pump beam is incident on PD3 while the reflected probe beam is incident on PD4.  The RF output of the pump photodetector is split with one output being demodulated to derive a PDH error signal that is then fed back to the cavity PZT mirror in order to lock the ring cavity to the laser frequency.  The other RF splitter output is then attenuated and subtracted from the RF probe photodetector output (PD4) prior to being demodulated.  This subtracted output can then be used to both detect weak atomic transitions as well as provide an error signal to lock the laser to a hyperfine transition.

\begin{figure}[htb]
  \begin{center}
  \includegraphics[width=8.5cm]{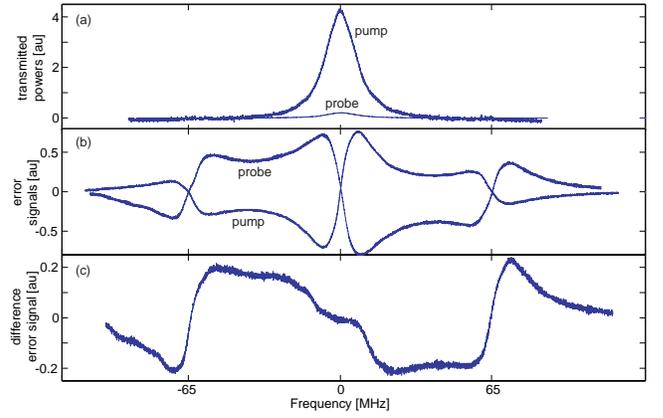}
  \end{center}
  \caption[TESTX]{Scanning bow-tie cavity mirror PZT with the Iodine cell removed. (a) Transmitted pump and probe powers, (b) Pump, Probe and (c) Difference PDH error signals.}
  \label{fig2}
\end{figure}

In order to balance the subtraction error signal, the iodine cell was removed from the cavity and the PZT mirror used to scan the cavity resonance across the frequency of the doubled laser output.  Figure \ref{fig2}a shows the transmitted powers of both the pump and probe beams.  Figure \ref{fig2}b shows the corresponding error signals at the subtracted output.  The pump error signal is recorded with the probe detector (PD4) blocked while the probe error signal is recorded with the pump detector (PD3) blocked.  After electronic balancing, using the variable attenuator, the subtracted error signal shown in figure \ref{fig2}c was obtained.  While trace \ref{fig2}c is clearly nonzero away from cavity resonance, the flat error signal through resonance indicates that when locked, this output is insensitive to cavity displacement.

\begin{figure}[htb]
  \begin{center}
  \includegraphics[width=8.5cm]{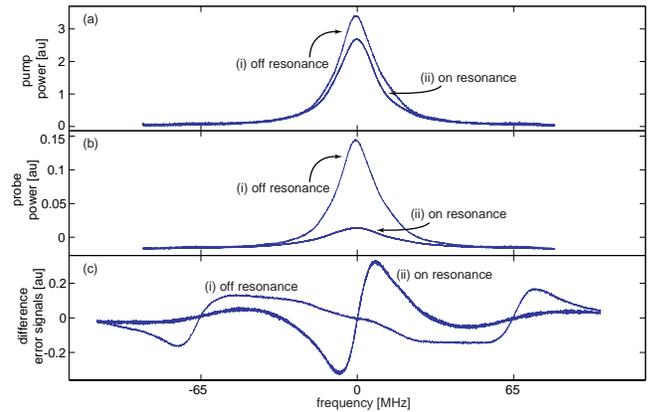}
  \end{center}
  \caption[TESTX]{Scanning FPI mirror PZT. Iodine cell inserted. (a) Pump transmitted powers with the laser frequency tuned both (i) off ($\approx$~6~GHz away) and (ii) on the centre of the R(56)32-0 broadened atomic transition, (b) shows the same for the probe, and (c) associated demodulated RF difference signals.}
  \label{fig3}
\end{figure}

After replacing the iodine cell, the FPI was once again scanned across the frequency of the doubled laser output. Traces \ref{fig3}a(i) and \ref{fig3}b(i) show the cavity transmitted power for the pump and probe respectively when the laser frequency is approximately 6GHz detuned from the R(56)32-0 iodine atomic transition.  In this regime the atomic absorption, both linear and non-linear, is negligible and the ratio of pump to probe transmitted power is approximately equal to the splitter ratio (19:1).  Traces \ref{fig3}a(ii) and \ref{fig3}b(ii) show the transmitted power of the pump and probe when the laser is tuned near the centre of the R(56)32-0 resonance.  Clearly the pump beam suffers only minor attenuation while the probe beam is nearly completely absorbed.   This differential bleaching of the pump/probe beams causes a significant difference in finesse seen by the pump and probe beams on reflection of the FPI.  The finesse difference unbalances the subtracted error signal generating the response shown by trace \ref{fig3}c(ii).  The error signal slope across resonance yields a direct measure of the differential non-linearity in the intra-cavity iodine cell. 

\begin{figure}[htb]
  \begin{center}
  \includegraphics[width=8.5cm]{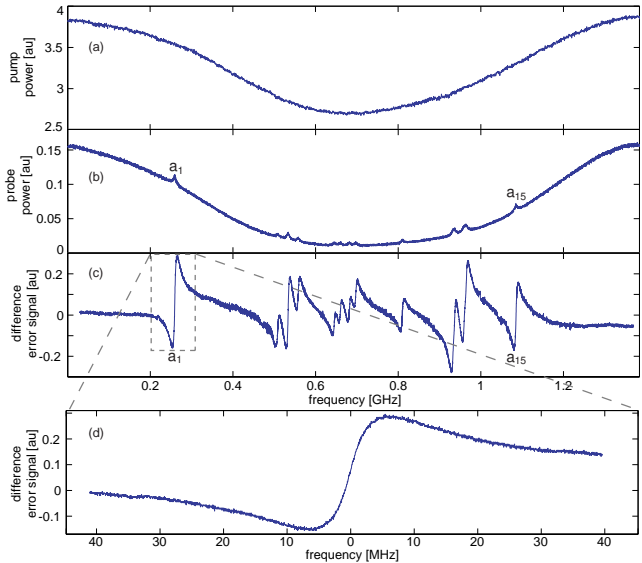}
  \end{center}
  \caption[TESTX]{FPI locked, broad laser frequency scan ($\approx$~1.4~GHz) across the R(56)32-0 broadened atomic resonance, showing absorption profile for (a) pump power, (b) probe power, (c) associated error signals from demodulated pump-probe difference, and (d) zoom in of R(56)32-0 a$_1$ hyperfine resonance difference error signal. All traces recorded with an iodine cell temperature of $\approx$~0$^{o}$ C. Traces (c) and (d) are taken with a measurement bandwidth of 1~ kHz.}
  \label{fig4}
\end{figure}

Figure \ref{fig4} shows the resulting traces when the FPI is locked on resonance with the pump beam and the laser frequency is then scanned across the Doppler broadened R(56)32-0 transition.  Traces \ref{fig4}a and \ref{fig4}b show the transmitted power of the pump and probe beams, respectively.  Due to cross saturation inside the iodine cell, the probe transmission exhibits inverted lamb dips at the hyperfine transition frequencies.  The cavity bounce number enhances the size of these inverted lamb dips showing clearly defined features.  Trace  \ref{fig4}c shows the output of the subtracted error signal.  At every hyperfine transition there is a clear error signal due to the differential phase response of the inverted lamb dips seen only by the probe beam. Trace \ref{fig4}c is recorded with a large intra-cavity pump power of approximately 71mW and shows substantial power broadening, leaving several hyperfine pairs unresolved.  Even so, the subtracted and demodulated output demonstrates a clear ability to lock the frequency of the laser to any of the well defined error signals.  Trace \ref{fig4}d shows a detailed scan of the R(56)32-0 a$_1$ hyperfine error signal, chosen for itÕs relative isolation, simplicity and large size. 

\begin{figure}[htb]
  \begin{center}
  \includegraphics[width=8.5cm]{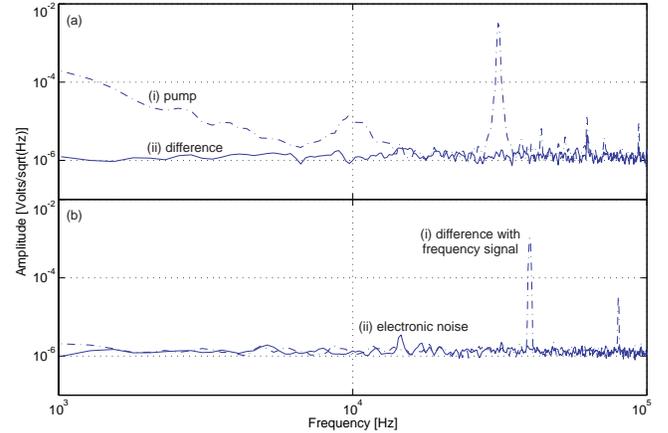}
  \end{center}
  \caption[TESTX]{Error signal spectra with FPI locked showing: a(i) pump only error signal, a(ii) pump-probe subtraction, canceling mechanical signal at $\approx$~30~kHz, b(i) showing broadband mechanical noise cancelation and frequency signal at $\approx$~40~kHz and b(ii) electronic noise floor.}  
  \label{fig5}
\end{figure}

With the ring cavity still locked and the laser frequency tuned to sit at the centre of the a$_1$ hyperfine resonance, we introduced a large 30kHz mechanical signal to the interferometer via the PZT mirror.  This produces a large signal clearly visible in the pump error signal spectrum of figure \ref{fig5}a(i).  The cavity locking servo for the pump PDH error signal has a unity gain bandwidth of approximately 7kHz, a simple 1/f response below unity gain frequency and a third order, low pass, elliptic filter above the unity gain frequency .  This ensures that the signal injected at 30kHz is free from servo effects and remains unsuppressed while FPI mechanical noise below 7kHz is suppressed by approximately the inverse of the loop gain. The subtracted error signal however, completely removes this 30kHz signal demonstrating mechanical noise cancellation of greater than 60dB (trace \ref{fig5}a(ii)).  In addition, this error signal is also free from the residual low frequency acoustic noise visible on the pump error signal spectrum.

Figure \ref{fig5}b demonstrates that when operating at the centre of the R(56)32-0 a$_1$ hyperfine resonance, the subtracted error signal responds to a deliberate laser frequency modulation at 40kHz even though it rejects mechanical noise; the introduced 30kHz mechanical signal is absent as is the low frequency residual noise.  Hence the locked cavity subtracted error signal discriminates between laser frequency noise and mechanical/acoustic noise, making this error signal an ideal optical frequency reference.

This research was completed under the auspices of the Centre for Gravitational Physics and Centre for Quantum Atom Optics supported by the Australian Research Council and the Australian Capital Territory Government.

\end{document}